\documentclass[12pt]{article}
 \usepackage[dvips]{graphicx}
 \usepackage[dvips]{graphics}
 \usepackage{amsmath}
 \usepackage{enumerate}
 \usepackage{psfrag}

  \usepackage[nomarkers,tablesfirst]{endfloat}

 \newcommand{\newc}{\newcommand}
 \newc{\ra}{\rightarrow}
 \newc{\lra}{\leftrightarrow}
 \newc{\beq}{\begin{equation}}
 \newc{\eeq}{\end{equation}}
 \newc{\barr}{\begin{eqnarray}}
 \newc{\barra}{\begin{eqnarray*}}
 \newc{\earr}{\end{eqnarray}}
 \newc{\earra}{\end{eqnarray*}}
 \newc{\texa}{\textstyle}
 \newc{\paral}{\parallel}
 \newc{\und}{\underline}
 \newc{\pars}{\partial}
 \newc{\nonu}{\nonumber \\}
 \newc{\nln}{\\ \vspace{2mm}}
 \newcommand{\Fc}{ {\mathcal{F}} }

\begin{document}

 \begin{center}
 {\large \bf On axisymmetric MHD equilibria with incompressible flows under side conditions}
 \vspace{3mm}

 {\large  G. N. Throumoulopoulos$^1$, H. Tasso$^2$, G.
 Poulipoulis$^1$} \vspace{2mm}

 {\it
 $^1$University of Ioannina, Association Euratom - Hellenic Republic,\\
 Section of Theoretical Physics, GR 451 10 Ioannina, Greece}
 \vspace{2mm}

 { \it  $^2$Max-Planck-Institut f\"{u}r Plasmaphysik, Euratom
 Association,\\
 D-85748 Garching, Germany }
 \end{center}
 %
 %
 \vspace{2mm}
 \begin{center}
 {\bf Abstract}
 \end{center}

 \noindent
 Axisymmetric equilibria with incompressible flows of
 arbitrary direction are studied in the framework of
 magnetohydrodynamics under a variety of physically relevant side
 conditions. To this end a set of pertinent non-linear ODEs  are
 transformed to quasilinear ones and the respective initial value
 problem is solved numerically with appropriately determined initial
 values near the magnetic axis. Several equilibria
 are then constructed surface by surface. The non field aligned flow
 results in novel configurations with a single magnetic axis,  toroidal shell
 configurations
  in which the plasma is confined
 within a couple of magnetic surfaces and double shell-like configurations.  In addition, the flow affects
 the elongation and triangularity of the magnetic surfaces.

 \newpage
 \begin{center}
 {\bf \large 1.\ Introduction}
 \end{center}

 In a previous paper \cite{TaTh1} the first two authors derived a generalized
 Grad-Shafranov equation governing the magnetohydrodynamic (MHD) equilibrium states
 of an axisymmetric plasma with incompressible flows [Eq. (\ref{1}) in section 2]. By assignment of the free
 functions contained in (\ref{1}), known
 solutions of the Grad-Shafranov equation, i.e the Solov\'ev solution \cite{Sh,So} and the
 Hernegger-Maschke solution \cite{He,Ma}, were extended in Refs. \cite{SiTh} and
 \cite{PoTh}, respectively. For both extended solutions   the
 flow can change the magnetic field topology, thus resulting
 in a variety of new configurations of astrophysical and laboratory
 concern.

 Instead of assigning the free surface functions  of
  (\ref{1}), it may be of physical or mathematical
importance to introduce side conditions, e.g. isodynamicity:
 $B^2=B^2(\psi)$,  where
 $\psi$ and $B$ are the poloidal magnetic flux function and the magnetic field modulus, respectively.
 In the quasi-static
  case,  viz. when the flow is neglected in the momentum
  equation but it is kept in Ohm's law,
 it was proved in
 Ref. \cite{Pa} that there is a unique configuration of this kind with
 circular magnetic surface cross-sections near the magnetic axis.
 This equilibrium was fully constructed in Ref. \cite{BiTa}. The same configuration
 persists in the
 case of flows aligned to the magnetic field  \cite{TaTh1}.
 In the case of
 non field aligned flows satisfying the side condition
 $P+B^2/2=f(\psi)$, where $P$ is the thermal pressure and $f$ is an arbitrary smooth
 function of $\psi$, the magnetic
 surfaces near  axis become elliptical with elongation
 perpendicular to the axis of symmetry \cite{Sch}.

 The aim of the present study is to construct axisymmetric steady states
 with incompressible flows of generic direction under a variety
 of side conditions and to examine the impact of the
 flow on the equilibrium characteristics and particularly in connection
 with the magnetic topology. A preliminary investigation  was conducted in Ref.
 \cite{TaTh2}. Here, the construction  is carried
 out numerically on the basis of a procedure suggested in Refs. \cite{Pa},
 \cite{BiTa} and \cite{TaTh1}. The main conclusion is that the flow results in a variety
 of novel equilibria and  opens
 up the possibility of changing the magnetic field topology.

 The side conditioned equilibrium equations are briefly reviewed in section
 2
 along with the solving procedure of Refs. \cite{Pa} and \cite{TaTh1}.
 In addition, the original  ODEs of concern are  mapped
 to quasi-linear ones by a transformation which for the quasi-static
 case was employed in Ref. \cite{BiTa}. After establishing initial values of the
 unknown functions near the magnetic axis the problem becomes well posed and is solved  numerically.
 The various kinds of configurations associated with numerical solutions  are presented
 in section 3 and are compared with the existing ones in the literature. Section 4 summarizes the study
 and the  conclusions.\\

 \noindent
 {\bf {\large 2. Side conditioned equilibria}}

\noindent
 The MHD equilibrium states of an axisymmetric  magnetized plasma
 with incompressible flows
 are determined by the generalized Grad-Shafranov equation
 \cite{TaTh1},
 \beq (1-M^2) \Delta^\star \psi -
 \frac{1}{2}(M^2)^\prime |\nabla \psi|^2
 + \frac{1}{2}\left(\frac{X^2}{1-M^2}\right)^\prime
 + R^2 P_s^\prime + \frac{R^4}{2}\left(\frac{\rho
 (\Phi^\prime)^2}{1-M^2}\right)^\prime = 0,
 \label{1}
 \eeq
 along with the Bernoulli relation for the pressure
 \beq
 P=P_s(\psi) - \rho\left\lbrack \frac{v^2}{2} - \frac{R^2
 (\Phi^\prime)^2}{1-M^2}\right\rbrack.
 \label{2}
 \eeq
 Here, $(z,R,\phi)$ are cylindrical coordinates with $z$
 corresponding to the axis of symmetry;  the
 function $\psi(R,z)$  labels the magnetic surfaces; $M(\psi)$ is
 the Mach function of the poloidal velocity with respect to the
 poloidal-magnetic-field Alfv\'en velocity;
 $\rho(\psi)$ and $\Phi(\psi)$ are  the density and  the electrostatic
 potential;
 $X(\psi)$ relates to the toroidal magnetic
 field;
 for vanishing flow the surface function $P_s(\psi)$
  coincides with the pressure; $v$ is the velocity
 modulus which can be expressed in terms of surface functions and $R$;  $\Delta^\star=R^2\nabla\cdot(\nabla/R^2)$;
  and the prime denotes a derivative with respect to $\psi$.
 Derivation of  (\ref{1}) and (\ref{2}) is
 provided  in Ref. \cite{TaTh1}.
  The surface quantities
$M(\psi)$, $\Phi(\psi)$, $X(\psi)$, $\rho (\psi)$ and $P_s(\psi)$
are free functions for each choice of which
 (\ref{1})  is fully determined and can be solved whence the boundary
condition for $\psi$ is given.

 By using the transformation
\begin{equation}
u(\psi) = \int_{0}^{\psi}\left\lbrack 1 -
M^{2}(g)\right\rbrack^{1/2} dg,
\end{equation}
 (\ref{1}) and (\ref{2}), respectively,  reduce to
 \beq
 \Delta^\star u
 + \frac{1}{2}\frac{d}{du}\left(\frac{X^2}{1-M^2}\right) + R^2\frac{d
 P_s}{d u}+\frac{R^4}{2} \frac{d}{du}
 \left(\rho \frac{d \Phi}{du}\right)^2  = 0,
 \label{3}
 \eeq
 \beq
 P=P_s(\psi) - \rho\left\lbrack \frac{v^2}{2} -
 R^2\left(\frac{d\Phi}{du}\right)^2 \right\rbrack.
 \label{3a1}
 \eeq
 Note that no quadratic term as $|{\bf\nabla}u|^{2}$ appears anymore
in (\ref{3}). The forms of  (\ref{3}) and (\ref{3a1}) indicate one
to introduce the new surface quantities
 \beq
 N(u)=\frac{X}{\sqrt{1-M^2}}, \ \ \
 L(u)=\sqrt{\rho}\frac{d\Phi}{d u},
 \label{3d}
 \eeq
which are helpful in reducing the number of the explicit free
functions
 by one.

 Instead of specifying the free functions  to
determine  (\ref{1}), one can introduce  side conditions as those
quoted in table \ref{table:1}.
 They can be expressed in terms of the
thermal pressure $P$, the magnetic pressure  $B^2/2$, the flow
energy density $\rho v^2/2$ or combinations of them and consist in
that these quantities remain uniform on magnetic surfaces.
 \begin{table}[!h]
 \begin{center}
 \begin{tabular}{|c|c|c|}
  \hline
 Side condition  &  $j(u)$&  $k(u)$ \\
 \hline
 \rule{0pt}{4ex} $P=P(u)$ & $\frac{{\textstyle 2\left\lbrack LMN+\left(1-M^2\right)\left(P_s-P\right)
 \right\rbrack }}
 {{\textstyle M^2\left(1-M^2\right)}}$ &
 $\frac{{\textstyle L^2\left(1-2M^2\right)}}{{\textstyle
 M^2\left(1-M^2\right)}}$\\[0.1cm]
 \hline
 \rule{0pt}{4ex} $B^2=B^2(u)$ & $B^2+\frac{{\textstyle 2LMN}}{{\textstyle 1-M^2}}$&$-\frac{{\textstyle L^2
 M^2}}{{\textstyle 1-M^2}}$\\[0.1cm]
 \hline
 \rule{0pt}{4ex} $\rho v^2=f(u)$&$\frac{{\textstyle 1}}{{\textstyle M^2}}\left(\frac{{\textstyle 2 L M
 N}}{{\textstyle 1-M^2}}+f \right)$ &$-\frac{{\textstyle L^2}}{{\textstyle M^2\left(1-M^2\right)}}$\\[0.1cm]
 \hline
 \rule{0pt}{4ex} $P+B^2/2=f(u)$&$\frac{{\textstyle 2\left(f-P_s\right)}}{{\textstyle 1-M^2}}$ &
 $-\frac{{\textstyle L^2}}{{\textstyle1-M^2}}$  \\[0.1cm]
 \hline
 \rule{0pt}{4ex} $B^2+\rho v^2 =f(u)$& $\frac{{\textstyle 4 L M
 N+f\left(1-M^2\right)}}{{\textstyle 1-M^4}}$ &
 $-\frac{{\textstyle L^2}}{{\textstyle 1-M^2}}$ \\[0.1cm]
 \hline
 \rule{0pt}{4ex} $P+B^2/2+\rho v^2/2=f(u)$ & $\frac{{\textstyle 2 L M
 N +\left(1-M^2\right)\left(f-P_s\right)}}
 {{\textstyle 1-M^2}}$ & $\frac{{\textstyle L^2\left(M^2-2\right)}}{{\textstyle 1-M^2}}$ \\[0.1cm]
 \hline
 \end{tabular}
 \caption{The coefficients $j(u)$ and $k(u)$ for several side conditions
 involving the quantities $P$, $B^2$ and $\rho v^2$. Note that the condition
 $P+\rho v^2/2=f(u)$ not included in the Table can be satisfied
 only for parallel flows
 [$d\Phi/du=0]$ because of the explicit $R$-dependence of the last term in
 (\ref{3a1}).}
 \label{table:1}
 \end{center}
\end{table}
 Such  conditions lead, in general, to an additional relation
 between $(\nabla u)^{2}, u $ and $R$ as already accomplished in
 Ref. \cite{TaTh1}. Specifically, (\ref{3}) and (\ref{3a1}) can
 be put in the respective forms
 \beq
 |{\bf\nabla}u|^{2} = 2[i(u) + R^{2}j(u) + R^{4}k(u)]
 \label{3a}
 \eeq
 \beq
 \Delta^{*}u = - f(u) - R^{2}g(u) - R^{4}h(u)
 \label{3b}
 \eeq
 where
 \beq
 i(u)=-\frac{N(u)^2}{2},
 \label{3c1}
 \eeq
 \beq
 f(u) = \frac{1}{2}\frac{d N(u)^2}{du}=-\frac{d i(u)}{du},
 \label{4}
 \eeq
 \beq
 g(u) = \frac{dP_{s}(u)}{du},
 \label{5}
 \eeq
 \beq
 h(u) =
 \frac{1}{2}\frac{d L(u)^2}{du}.
 \label{6}
 \eeq
 The other coefficients $j(u)$ and $k(u)$ being side
 condition dependent, are given in table \ref{table:1}.

 Equations (\ref{3a}) and (\ref{3b}) can be solved  simultaneously
 by the method suggested by Palumbo \cite{Pa} according to which the variable
 $z$ is treated as function of $u$ and $R$. The method leads to
 compatibility conditions in the form of five ordinary differential
 equations (ODEs) [Eqs. (44)-(48) of Ref. \cite{TaTh1}] along with an integral relation for the function
 $z(u,R)$ [Eq. (49) of Ref. \cite{TaTh1}]. Thus, to determine the equilibrium a) the  ODEs should be solved on each
 magnetic surface and then b) the  intersection points of that surface
 with the poloidal cross-section should  be obtained by the integral relation.
  There is a difficulty, however,   stemming from  the non linearity of
 the ODEs. Here, to overcome this difficulty, the function $u$ is mapped to a new
 function
  $w$ through the transformation:
 \beq
 \frac{du}{dw}=-\left(g+\frac{dj}{du}\right)\equiv \Fc(w),
 \label{16}
 \eeq
 where $\Fc(w)$ is an arbitrary smooth function.

 Under this transformation the compatibility condition leads to the
 following set of ODEs
 \beq
 4Q+2\Theta Y^\prime-Y(Y+\Theta^\prime)=0,
 \label{21}
 \eeq
 \beq
 8\Xi+Y(4-Q^\prime)+2Q Y^\prime +\Theta^\prime=0,
 \label{22}
 \eeq
 \beq
 -3-2\Theta H^\prime +Q^\prime +2\Xi Y^\prime +H(6 Y+\Theta^\prime -Y\Xi^\prime)=0,
 \label{23}
 \eeq
 \beq
 -2QH^\prime+H(Q^\prime-8) + \Xi^\prime=0,
 \label{24}
 \eeq
 \beq
 -5H^2-2\Xi H^\prime + H\Xi^\prime=0,
 \label{25}
 \eeq
 where the functions $Q(w)$, $\Theta(w)$, $Y(w)$, $\Xi(w)$ and $H(w)$ can be expressed in terms of the initial physical
 surface quantities.
  Note that (\ref{21})-(\ref{25}) are {\em quasi-linear}, viz. the derivatives appear
  linearly, and therefore  Picard' s theorem guarantees existence and
  uniqueness of the respective initial value problem.  Because of indefiniteness of some
  of   (\ref{21})-(\ref{25}) on the magnetic axis when they are put in solved
  forms, initial
  values of the unknown functions near the magnetic axis can be obtained on the basis of
  Mercier expansions around axis along with a l'Hospital-like procedure.
  Details will be given elsewhere. Also, the integral
  relation for $z(w,x)$ is given by
   \beq
 \left.\frac{\partial z}{\partial x}\right|_{w}=
 -\frac{p}{q} = \frac{\pm\frac{1}{4}[Hx^2 + x-Y]}
 {\left\lbrack \Theta +Q x +x^2 \Xi -\frac{1}{4} x \left(H x^2 +x-
 Y\right)^2
 \right\rbrack^{1/2}},
 \label{20}
 \eeq
where $x=R^2$, $p\equiv\pars w/\pars x$, $q\equiv\pars w/\pars z$,
$r\equiv\pars^2w/\pars x^2$ and
 $q\equiv\pars^2w/\pars z^2$.
 Once initial values are established the problem is well posed and can be solved numerically.
 Accordingly, we have
  developed a programme in  Mathematica 5.1.
 For up-down symmetric configurations to
  be considered here
  there are three free parameters ($R_0$, $\Xi_0$ and $H_0$)
  associated with
  the radial distance of the magnetic axis  and the
   functions $\Xi$ and $H$ thereon.  It is also noted that  the problem can be solved by the transformation,
  alternative to (\ref{16})
  \beq
 \frac{du}{d w}=-\frac{1}{2}\left(h+\frac{d k}{du}\right)\equiv \Fc(w).
 \label{40}
 \eeq
  \\

 \noindent
 {\bf {\large 3. Magnetic configurations and impact of flow}}

 For the solutions to be presented here we have  chosen, without
 loss of generality,  $R_0=1$ and  $w_0=0$, where $w_0$ is the value
  of $w$ on axis. Depending on the sign of
   $\Xi_0$, there are two kinds of configurations:

 \begin{enumerate}

 \item
Toroidal configurations for $\Xi_0<0$ having  a single  magnetic
axis
   [located on $(z=0, R=R_0)$ where $w=w_0=0$]. These configurations have  magnetic  field topology
  similar to those of Refs. \cite{BiTa} and \cite{Sch}.
  $w$-contours  of such a solution are shown in  Fig.
  \ref{fig:1}.
  In general for $\Xi_0<0$ the magnetic surfaces are
   more elongated horizontally
  (parallel to the mid-plane $z=0$) up to the magnetic axis as compared with the quasi-static
  ones corresponding to $\Xi_0=H_0=0$.

  \begin{figure}[!h]
 \begin{center}
 \psfrag{z}{$z$}
 \psfrag{R}{$R$}
 \includegraphics[scale=0.8]{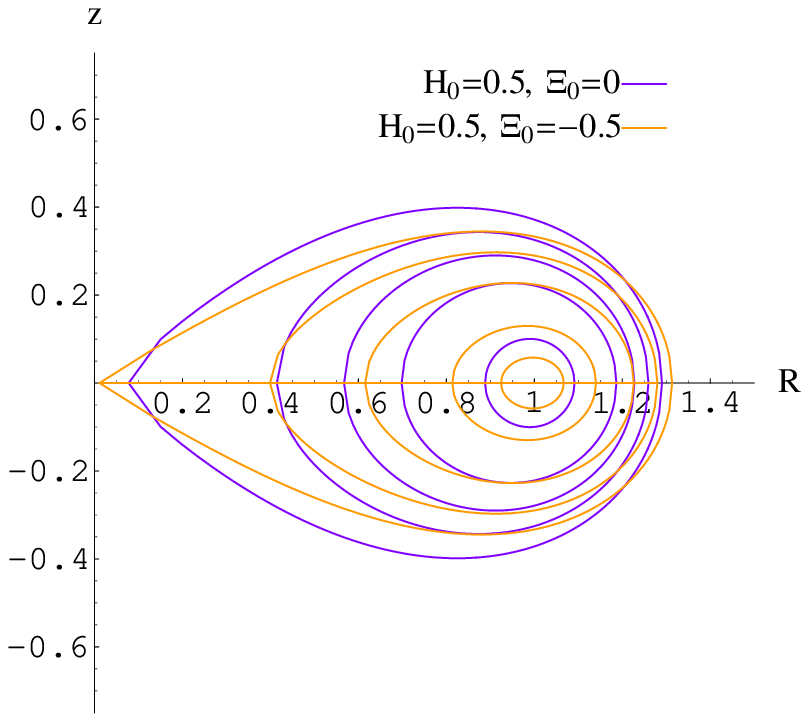}
 \caption{Two sets of  $w$-contours the one (in yellow) for  $\Xi_0=-0.5$
 and $H_0=0.5$ and the other (in purple) for  $\Xi_0=0$ and $H_0=0.5$. Both
 equilibria have a magnetic axis on which $w=0$.}
 \label{fig:1}
 \end{center}
\end{figure}

 \item
  Toroidal shells for $\Xi_0>0$ in which
  the plasma is contained within two toroidal surfaces. $w$-contours for a solution
  of this kind are shown in Fig. \ref{fig:2}. Although mathematically the solution has an extremum
  on $w=0$, the
  physically acceptable part  is restricted to non
  positive values of $\Theta(w)$,  thus resulting in  a toroidal
  vertically elongated shell.


 \begin{figure}[!h]
 \begin{center}
 \psfrag{z}{$z$}
 \psfrag{R}{$R$}
 \includegraphics[scale=0.8]{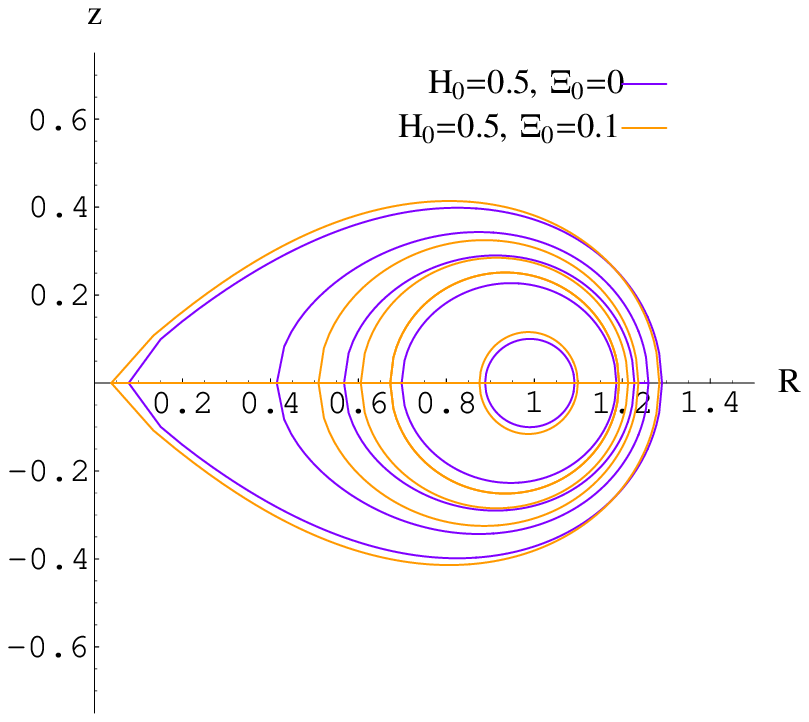}
 \caption{A set of $w$-contours associated with a toroidal
 shell for  $\Xi_0=0.1$ and  $H_0=0.5$ (in yellow).
 The second   set of $w$-contours (in purple) corresponds to an
 equilibrium with a magnetic axis ($\Xi_0=0$ and $H_0=0.5$).}
 \label{fig:2}
 \end{center}
\end{figure}

 \end{enumerate}

 The various
  kinds of configurations can also be classified in terms of the
  parameters $\Xi_0$ and $H_0$ as follows.
 \begin{enumerate}
  \item $\Xi_0=0$ and $H_0\neq 0$:\\
 Let us first note that $\Xi(w)=0$ implies parallel  flows (or the quasi-static equilibrium)
  because then  it follows from
  (\ref{25}) that $H(w)=0$. This should not be confused with the case of $\Xi_0=0$ (on axis) in this paragraph
   which involves
  non parallel flows when $H_0\neq 0$.
  As in the  case of parallel flows, however,
  the magnetic surfaces near axis have circular cross sections.
  For $H_0>0$ the  surfaces far from axis are less parallel  elongated
  than  those of the quasi-static
  equilibrium. For $H<0$ the triangularity can change drastically. As an
  example,
  a  configuration with inverse triangularity is shown in Fig.
  \ref{fig:3}.

  \begin{figure}[!h]
 \begin{center}
 \psfrag{z}{$z$}
 \psfrag{R}{$R$}
 \includegraphics[scale=0.8]{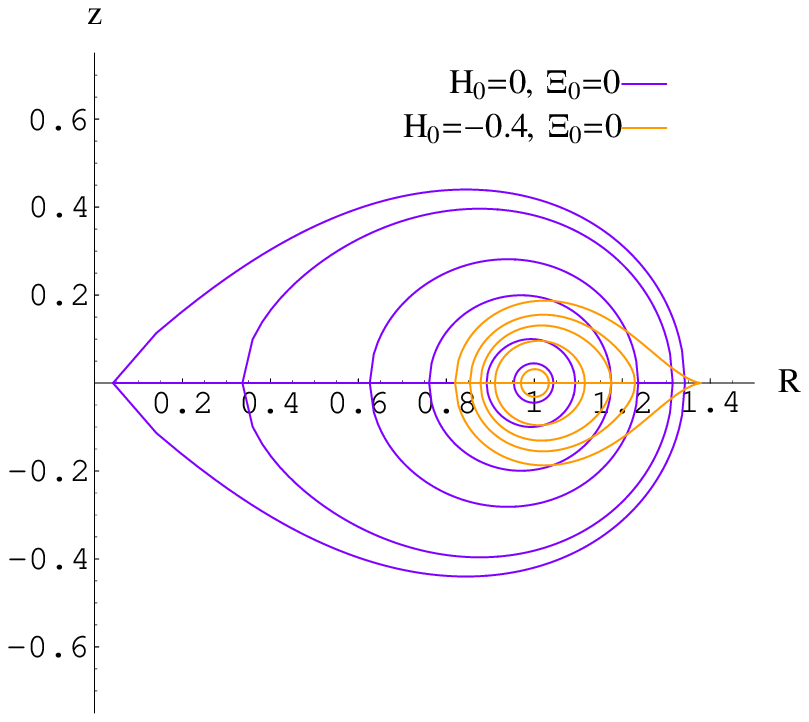}
 \caption{A configuration with inverse triangularity (in yellow) for $\Xi_0=0$ and
 $H_0=-0.4$. The $w$-curves in purple correspond to the quasi-static
 configuration.}
 \label{fig:3}
 \end{center}
\end{figure}

 \item $\Xi_0\neq 0$ and $H_0=0$:\\
 Equilibria for $\Xi_0<0$ and $H_0=0$  have been constructed in
  Ref. \cite{Sch}  by a different method.
  Unlike in Ref. \cite{Sch}, however, no restriction on the
  elongation of the magnetic surfaces (parallel to the mid-plane $z=0$) was found here.
  As a matter of fact
  such a configuration  with very elongated surfaces is
  presented in Fig. \ref{fig:4} (in purple). Thus, the limitation on the elongation
  reported in \cite{Sch} may be due to the particular  method
  of solution  in that paper.
  For $\Xi_0>0$ the equilibrium becomes a toroidal shell
   with  magnetic surfaces elongated perpendicular to the mid-plane $z=0$.

 \begin{figure}[!h]
 \begin{center}
 \psfrag{z}{$z$}
 \psfrag{R}{$R$}
 \includegraphics[scale=0.8]{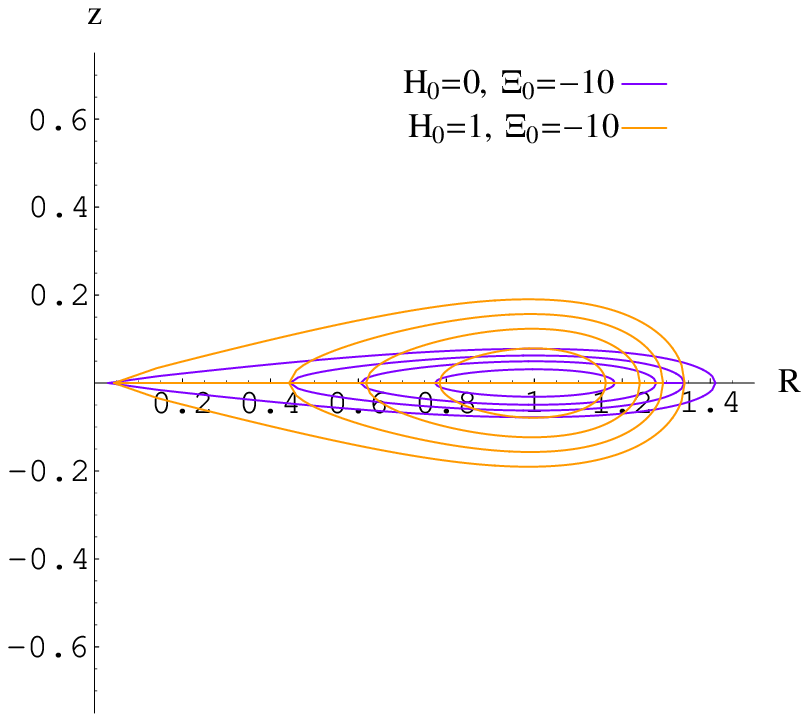}
 \caption{Two configurations with a magnetic axis and very elongated magnetic surfaces parallel to the
 mid-plane $z=0$, the one for $\Xi_0=-10$ and $H_0=0$ (in purple) and the other for $\Xi_0=-10$ and $H_0=1$
 (in yellow). The respective values of the eccentricity
 are 0.988 and 0.9.}
 \label{fig:4}
 \end{center}
\end{figure}

 \item $\Xi_0\neq 0$ and $H_0\neq 0$:\\
   In this generic case it is
  particularly interesting to examine whether there are
  configurations with two magnetic axes. This requires  two roots in the numerator of (\ref{20}) and four roots
   in the denominator appropriately
  located with respect to the roots of the numerator. (Note that for a quasi-static equilibrium ($H=\Xi=0$)
  only configurations with a single magnetic axis are possible.) For this
  reason we first examined this requirement by applying  the Sturm
  theorem and Descartes rule \cite{Bro}. The former determines the
  exact number of real roots of a polynomial with real coefficients;
   the latter determines the maximum number of
  positive roots of such a polynomial. It turns
  out that the requirement is compatible with the  Sturm
  theorem and Descartes  rule. Then, by inspection we found that the
  requirement can be fulfilled for $\Xi_0>0$ and $H_0<0$. An
  equilibrium of this kind shown in Fig. \ref{fig:5} consists of a
  toroidal shell    reaching the axis of symmetry, similar to
  those reported  in paragraph 3.2, and a second thin
  shell-like configuration located farther from the axis of symmetry. The distance between the
  two configurations decreases as $|\Xi_0/H_0|$ takes larger
  values. As can be seen in Fig. \ref{fig:5}, however,  the magnetic surfaces
  of the  shell-like configuration do not close. Closeness  does not improve
  either by varying $\Xi_0$ and $H_0$ or
  by using a different numerical method in FORTRAN. Therefore, the existence of double toroidal shell
  configurations with closed magnetic surfaces remains an open question.

 \begin{figure}[!h]
 \begin{center}
 \psfrag{z}{$z$}
 \psfrag{R}{$R$}
 \includegraphics[scale=0.8]{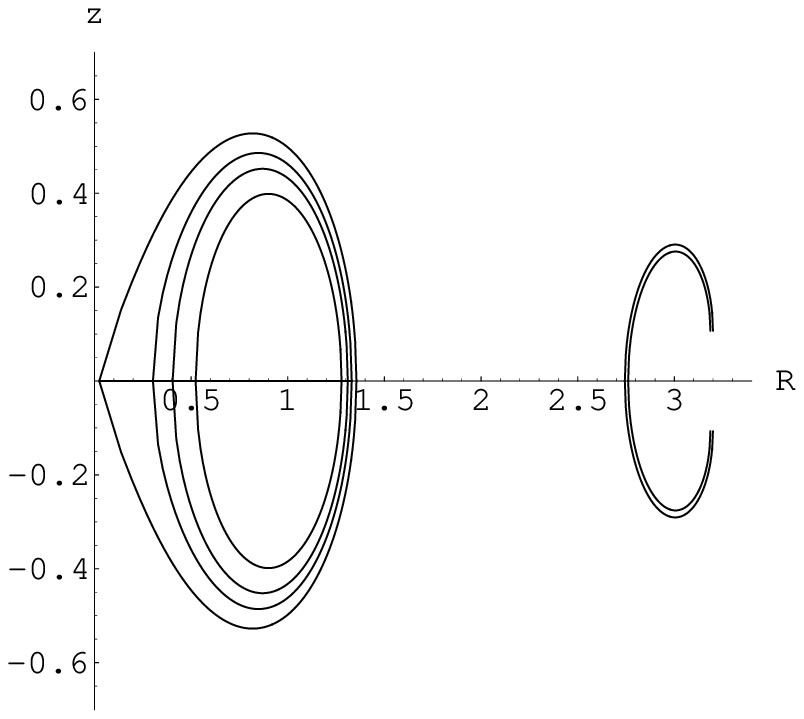}
 \caption{An equilibrium for $\Xi_0=0.05$ and $H_0=-0.1$ consisting of a toroidal shell and a smaller
 toroidal shell-like configuration.}
 \label{fig:5}
 \end{center}
\end{figure}

 For the other three combinations of signs of $\Xi_0$ and $H_0$ one
 can obtain configurations similar to those presented  in paragraph 3.2  As in the case of $H_0=0$, for $\Xi_0<0$
  there is no
 limitation on the elongation of the
  magnetic surfaces parallel to the mid-plane $z=0$  as $|\Xi_0|$ increases.
  Also, it is
 worth to mention the strong change in the triangularity of the
 configuration for $H_0<0$ as it is illustrated in Fig. \ref{fig:6}.

 \begin{figure}[!h]
 \begin{center}
 \psfrag{z}{$z$}
 \psfrag{R}{$R$}
 \includegraphics[scale=0.8]{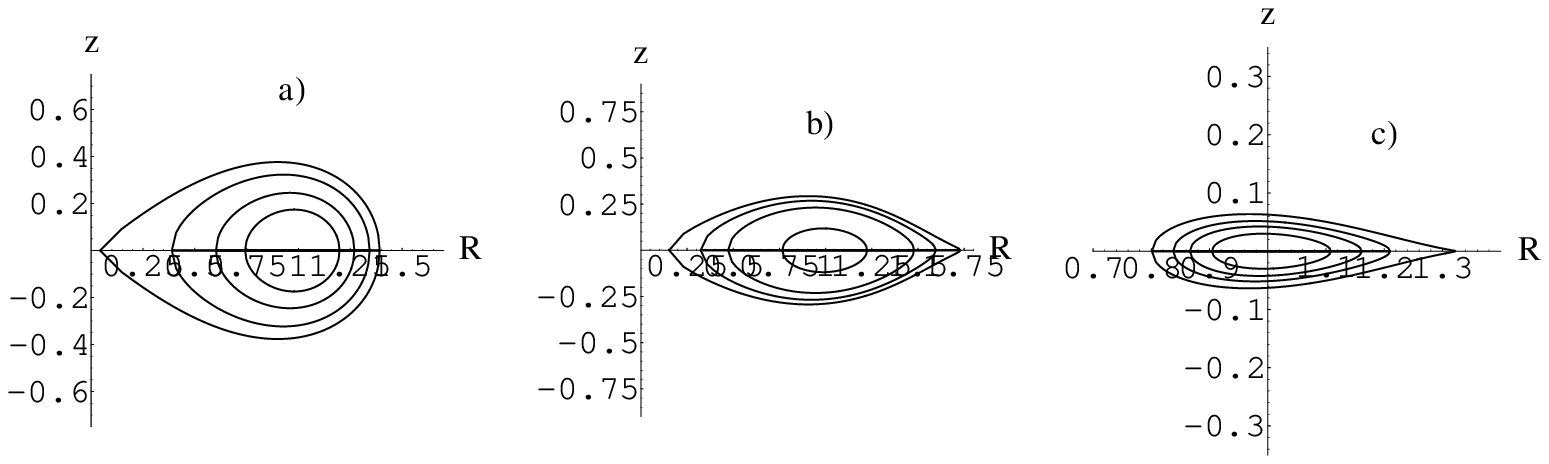}
 \caption{Impact of $H_0<0$ on the triangularity of an equilibrium with a magnetic axis ($\Xi_0=-0.1$):
 a) a quasi-static like configuration ($H_0=-0.09$);  b) an eye-like configuration ($H_0=-0.3$); and c) a configuration
 with inverse triangularity ($H_0=-0.4$).}
 \label{fig:6}
 \end{center}
\end{figure}

 \end{enumerate}

 {\bf {\large 4. Summary and Conclusions}}

We have studied axisymmetric equilibria with incompressible flows
 under side conditions of physical relevance by a procedure
 introduced in Refs. \cite{Pa}, \cite{BiTa} and \cite{TaTh1}. This procedure reduces the problem to a set of
 ODEs for certain surface functions and an integral
 relation determining the points of the cross section  of a magnetic surface  with the
 poloidal plane. Because of the nonlinearity of the original ODEs, we have employed
   transformation  (\ref{16}) mapping
 the original ODEs to quasilinear ones [Eqs. (\ref{21})-(\ref{25})] and containing equal
 number of unknown surface functions;
 thus,  existence and uniqueness of
 the respective initial value problem is guaranteed. After determining appropriately initial
 values near axis,  because of
 indefiniteness of the ODEs thereon, the
 problem  has been solved numerically surface by surface with
 two free parameters ($\Xi_0$ and $H_0$) associated with the non-
 field aligned flow.

 The  flow results in
 the following novel kinds of up-down symmetric equilibria:
 \begin{enumerate}
 \item
 Configurations with  a single magnetic axis for $\Xi_0\leq 0$ with  magnetic surfaces near axis elongated parallel
 to the mid-plane $z=0$. For $\Xi_0=0$  the  magnetic surfaces near axis
 become circular
  as the quasi-static ones corresponding to $\Xi_0=H_0=0$. The special case
 of equilibria  constructed
 by a different method in Ref. \cite{Sch} are recovered for $H_0=0$.
 Unlike in Ref. \cite{Sch}, however, no restriction on the
  elongation  has been
 found in the present study.
 \item Toroidal shells for   $\Xi_0>0$ and $H_0\geq 0$ in which the
 plasma is confined in the interior of two nested magnetic
 surfaces. The magnetic surfaces are  elongated perpendicular to the
 mid-plane $z=0$ compared with the quasi-static ones.
 \item Equilibria consisting of a toroidal shell reaching the
 axis of symmetry and a second shell-like configuration
 for $\Xi_0>0$ and $H_0<0$. Thus, the flow opens
 up the possibility of changing the magnetic field topology.
 \end{enumerate}
 Also, the shape of the magnetic surfaces far from axis is affected
 by the value
 and sign of $H_0$; specifically: a) they become less
 elongated parallel to the mid-plane $z=0$ as $H_0$ takes larger
 positive values and b) the triangularity of those surfaces is affected
 drastically for negative values of $H_0$.

 It is emphasized that  the above reported conclusions hold irrespective of the
 particular condition of table \ref{table:1} except for $P+B^2/2$
 being uniform on surfaces which corresponds to $H_0=0$. The
 properties of particular equilibria
 in connection with profiles of the (original) physical quantities, i.e. pressure, magnetic
 field,
 velocity etc, deserves further investigation. Also, in view of the tough and  in general  unsolved  stability problem
 of steady states with flow, the stability of the equilibria
 constructed here remains an open question.


 \begin{center}
 {\bf {\large Acknowledgements}}
 \end{center}

 Part of this work was conducted during a visit of  the first and third authors
 to the Max-Planck-Institut f\"{u}r Plasmaphysik, Garching.
 The hospitality of that Institute is greatly appreciated.

 The present work was performed under the Contract of Association
 ERB 5005 CT 99 0100 between the European Atomic Energy Community and
 the Hellenic Republic. The views and opinions expressed herein do
 not necessarily reflect those of the European Commission.

 \newpage
 \listoftables

 \newpage

 \listoffigures

 \end{document}